\documentclass{aa}  
\usepackage{graphicx}
\usepackage{txfonts}
\usepackage{natbib,twoopt}
\bibpunct{(}{)}{;}{a}{}{,}             
\usepackage{hyperref}
\hypersetup{
    colorlinks=true,
    linkcolor=blue,
    citecolor = blue,
    filecolor=magenta,      
    urlcolor=blue}

\authorrunning{Fanelli et al.}
\titlerunning{Terzan~6 chemical abundances}

\begin{document}

\title{Detailed chemical abundances of the globular cluster Terzan~6 in the inner bulge
\thanks{Based on observations collected at the W. M. Keck Observatory, which is operated as a scientific partnership among the California Institute of Technology, the University of California, and the National Aeronautics and Space Administration. The Observatory was made possible by the generous financial support of the W. M. Keck Foundation. \\ Also based on observations at the Very Large Telescope of the European Southern Observatory at Cerro Paranal (Chile) under program 089.D-0306 (PI:Ferraro).}}

\author{C. Fanelli\inst{1}, 
    L. Origlia\inst{1},    
    A. Mucciarelli\inst{2}\fnmsep\inst{1},
    F. R. Ferraro\inst{2}\fnmsep\inst{1},
    R. M. Rich\inst{3},
    B. Lanzoni\inst{2}\fnmsep\inst{1},
    D. Massari\inst{1},
    C. Pallanca\inst{2}\fnmsep\inst{1},
    E. Dalessandro\inst{1},
    and  M. Loriga\inst{2}\fnmsep\inst{1}
}

\institute{INAF, Osservatorio di Astrofisica e Scienza dello Spazio di Bologna, Via Gobetti 93/3, I-40129 Bologna, Italy\\
\email{cristiano.fanelli@inaf.it}
         \and
         Dipartimento di Fisica e Astronomia, Università degli Studi di Bologna, Via Gobetti 93/2, I-40129 Bologna, Italy
         \and
         Department of Physics and Astronomy, UCLA, 430 Portola Plaza, Box 951547, Los Angeles, CA 90095-1547, USA
         }
\abstract {
We used near-infrared spectroscopy at medium-high resolution (R=8,000$-$25,000) to perform the first comprehensive chemical study of the intermediate luminosity bulge globular cluster Terzan~6. 
We derived detailed abundances and abundance patterns of 27 giant stars, likely members of Terzan~6, based on their accurate Hubble Space Telescope proper motions and line-of-sight radial velocities. 
From the spectral analysis of these stars, we determined an average heliocentric radial velocity of 143.3$\pm$1.0 km s$^{-1}$ with a velocity dispersion of 5.1$\pm$0.7 km s$^{-1}$ and an average [Fe/H]=$-0.65\pm0.01$ and a low 1$\sigma$ dispersion of 0.03 dex.
We also measured some depletion of [Mn/Fe] with respect to the solar-scaled values and 
enhancement of for [Ca/Fe], [Si/Fe], [Mg/Fe], [Ti/Fe], [O/Fe], [Al/Fe], [Na/Fe], and, to a lower extent, for [K/Fe], consistent with previous measurements of other bulge globular clusters 
and favoring the scenario of a rapid bulge formation and chemical enrichment.
Some spread in the light element abundances suggest the presence of first- and second-generation stars, typical of genuine globulars.
Finally, we measured some depletion of carbon and low $\rm ^{12}C/^{13}C$ isotopic ratios, 
as in previous studies of field and cluster bulge giants, indicating that extra-mixing mechanisms 
should be at work during the post main sequence evolution in the high metallicity regime as well.}
\maketitle
\section{Introduction}
The study presented in this paper is part of a long-term, ongoing project aimed at studying the stellar populations of bulge GCs. By means of ground-based and Hubble Space Telescope (HST) NIR photometry and high-resolution spectroscopy from Keck and ESO telescopes, we have already characterized the photometric and chemical properties of the stellar population in several clusters \citep{origlia02,origlia04,origlia05,origlia08,valenti07,valenti10,valenti11,valenti15,saracino15,saracino16,saracino19,pallanca19,pallanca21a,pallanca23,cadelano20,cadelano23,deras23,leanza23}), unveiling the true nature of a few peculiar stellar systems, namely, Terzan~5 \citep{ferraro09,ferraro16,lanzoni10,origlia11,origlia13,origlia19,massari14b,massari15} and Liller~1 \citep{ferraro21,pallanca21b,dalessandro22,crociati23,deimer24}.  

The chemical characterization of these stellar systems is generally based on the spectral analysis of samples of luminous giant stars, which can be spectroscopically observed at medium-high resolution in the NIR within reasonable exposure times, thus maximizing the number of sampled targets.
By using the same observational strategy and spectral analysis, we have derived complementary chemical information  
of several tens of M giants in Baade’s window \citep{rich05} in three inner bulge fields located at (l,b) = ($0,-1$) \citep{rich07} and (l,b)=($0,-1.75$),($1,-2.65$) \citep{rich12}, and in the region surrounding the stellar system Terzan~5 \citep{massari14a}.

In this framework, the present paper is focused on Terzan 6, a highly reddened GC in the inner bulge of the Milky Way with an average color excess of $E(B-V)=2.35$, a galactocentric distance of just 1.3 kpc, and a distance from Earth of $\sim 7$ kpc (\citealp{valenti07}; see also \citealp{fahlman95, barbuy97, baumg_vasil21}). It has an intermediate luminosity ($M_V=-7.56$; \citealp{harris96}), an estimated total mass of $\sim 10^5 M_\odot$ \citep{baumgardt18}, and a very concentrated structure, suggesting that it is a post-core collapse system \citep{trager95}. 
\begin{figure*}[h]
    \centering
    \includegraphics[width=0.95\textwidth]{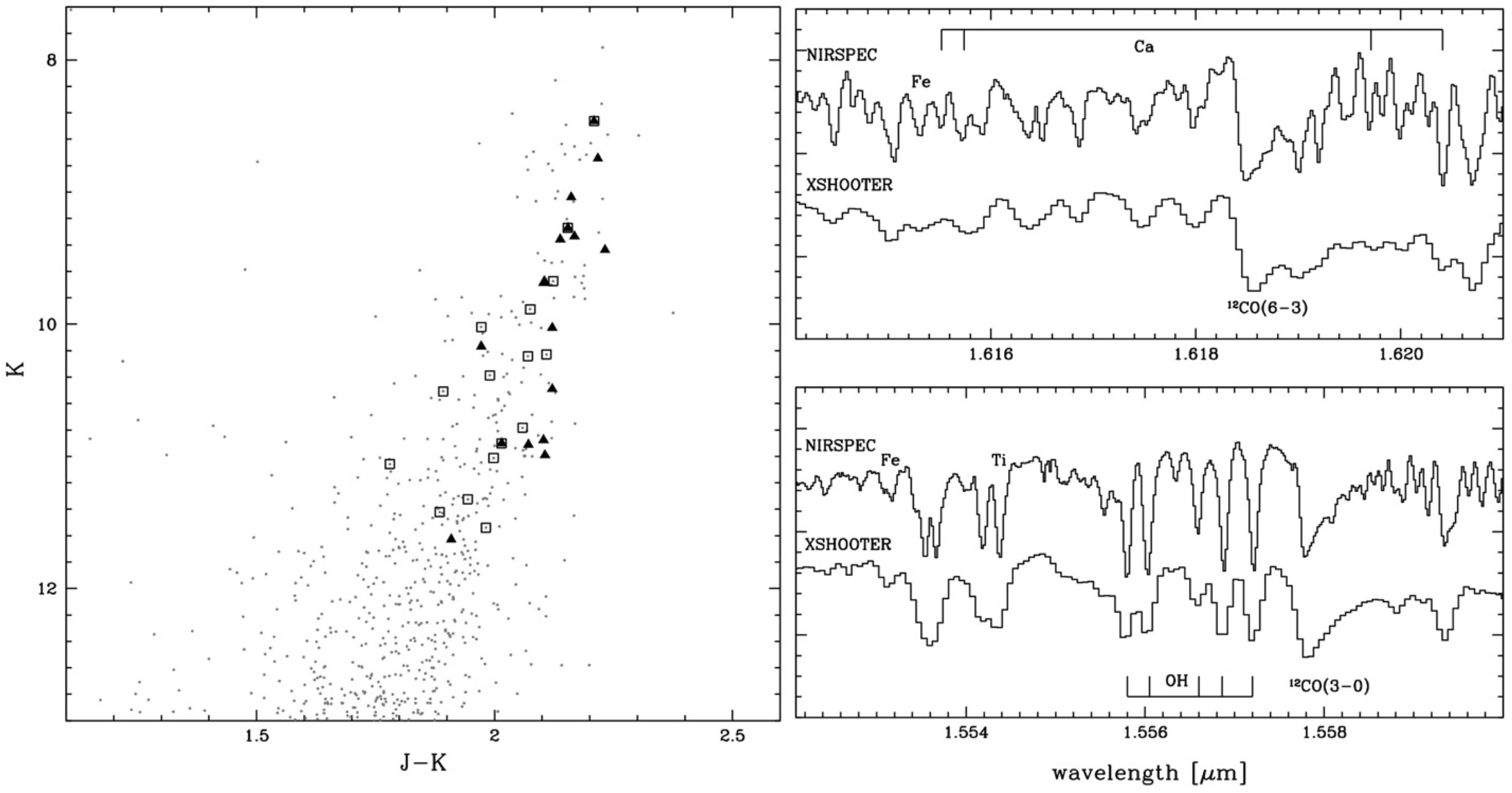}
    \caption{Photometric properties of the stars observed with NIRSpec and X-shooter. Left panel: Color-magnitude diagram of K,J-K (left panel) for Terzan~6. We have highlighted the stars observed with NIRSpec (empty squares) and with X-shooter (filled triangles). Right panel: Portions of the H-band spectra of star \#5 as observed with both X-shooter (at a resolution R$\approx$8,000) and NIRSpec (at R$\approx$25,000). Some features of interest for chemical abundance analysis are also marked in the panels.}
    \label{obs_stars}
\end{figure*}
Terzan 6 is mostly known for hosting a low-mass X-ray binary showing Type-I X-ray bursts, with sharp eclipses of the signal during the outburst phases (see \citealp{painter24} and references therein). As far as we know, the only estimates of the cluster metallicity published so far are those from a Ca triplet integrated light ([Fe/H]$=-0.65$ dex) by \citet{az88} and the photometric one ([Fe/H]=$-0.62$ dex) by \citet{valenti07}.
\begin{table}[!h]
\caption{Observed stars in Terzan~6.}
\label{tab1}
\scriptsize
\setlength{\tabcolsep}{5.3pt}
\renewcommand{\arraystretch}{1.2}
\begin{tabular}{|c|c|c|c|c|c|c|}
\hline\hline 
ID &  J & H & K & RA & Dec  & Instrument$^1$\\
 &[mag] & [mag] & [mag] & [Deg] & [Deg] &  \\ 
\hline
       5 &  10.67 & 9.18  &  8.46 & 267.7016903 & -31.2682732  &N,X \\
       9 &  10.96 &  9.48 &  8.74 & 267.7033392 & -31.2742385  &X \\
      10 &  11.20 &  9.69 &  9.04 & 267.6796299 & -31.2857720  &X \\
      12 &  11.42 &  9.88 &  9.27 & 267.6970891 & -31.2667239  &N,X \\
      13 &  11.50 & 10.00 &  9.33 & 267.7066404 & -31.2814706  &X \\
      14 &  11.50 & 10.01 &  9.36 & 267.6928178 & -31.2609826  &X \\
      18 &  11.67 & 10.12 &  9.44 & 267.7007463 & -31.2914683  &X \\
      20 &  11.78 & 10.27 &  9.67 & 267.6810107 & -31.2878000  &X \\
      21 &  11.79 & 10.31 &  9.69 & 267.6977345 & -31.2955471  &X \\
      22 &  11.80 & 10.28 &  9.67 & 267.7028318 & -31.2711766  &N \\
      28 &  11.96 & 10.48 &  9.89 & 267.6978652 & -31.2707428  &N \\
      31 &  12.15 & 10.63 & 10.03 & 267.7088901 & -31.2742202  &X \\
      32 &  11.99 & 10.56 & 10.02 & 267.6963902 & -31.2746828  &N \\
      33 &  12.14 & 10.72 & 10.17 & 267.6986477 & -31.2948039  &X \\
      35 &  12.31 & 10.85 & 10.24 & 267.6980300 & -31.2770139  &N \\
      37 &  12.34 & 10.81 & 10.23 & 267.6873144 & -31.2780827  &N \\
      39 &  12.38 & 10.96 & 10.39 & 267.6942027 & -31.2720726  &N \\
      46 &  12.40 & 11.07 & 10.51 & 267.6950460 & -31.2762939  &N \\
      48 &  12.61 & 11.07 & 10.49 & 267.7086115 & -31.2768451  &X \\
      65 &  12.84 & 11.31 & 10.78 & 267.6995094 & -31.2777294  &N \\
      67 &  12.92 & 11.48 & 10.90 & 267.6899446 & -31.2772069  &N,X \\
      69 &  12.84 & 11.56 & 11.06 & 267.6964627 & -31.2732291  &N \\
      70 &  12.98 & 11.47 & 10.88 & 267.6989533 & -31.2918849  &X \\
      74 &  12.98 & 11.49 & 10.91 & 267.6925713 & -31.2851530  &X \\
      82 &  13.10 & 11.58 & 10.99 & 267.6995912 & -31.2816276  &X \\
      83 &  13.01 & 11.58 & 11.01 & 267.6926224 & -31.2748025  &N \\
      90 &  13.27 & 11.86 & 11.32 & 267.6967447 & -31.2730136  &N \\
     101 &  13.30 & 11.94 & 11.42 & 267.6981256 & -31.2752570  &N \\
     114 &  13.52 & 12.09 & 11.54 & 267.6932534 & -31.2781304  &N \\
     116 &  13.54 & 12.17 & 11.63 & 267.6991636 & -31.2728558  &X \\
\hline\hline
\end{tabular}
\vspace{2pt}
\hspace{1pt}
$^1$~ Instrument with which the star has been observed: N=NIRSpec, X=X-shooter.
\end{table}While the photometric properties and the structural parameters of Terzan 6 will be discussed in a forthcoming paper (Loriga et al., in prep.), here we present the first, comprehensive chemical screening of Terzan~6 stellar population, based on medium-high resolution spectroscopy in the NIR of a representative sample of 27 giant stars, likely members of the cluster, as inferred by their 3D kinematics.
Section~\ref{obs} describes the spectroscopic observations and data reduction, Sect.\ref{memb} describes the derivation of the  proper motions (PMs) and radial velocities (RVs) of the observed stars, while their stellar parameters and chemical abundances are presented in Sect.~\ref{param} and \ref{abun}, respectively. Section~\ref{conc} presents some discussion of the obtained results in the context of the bulge GC population.

\section{Observations and data reduction}
\label{obs}

We used the NIR photometric catalog of Terzan~6 by \citet{valenti07} to compute the K,J-K color-magnitude diagram of Fig.~\ref{obs_stars} (left panel) and select suitable luminous giants for spectroscopic observations. 
We observed 16 luminous giants with NIRSpec  \citep{mclean98} at Keck II on 14 and 15 June 2012. We used the NIRSpec-5 setting to enable observations in the H-band and a 0.43" slit width that provided an overall spectral resolution R=25,000.
We observed 17 luminous giants with X-shooter \citep{vernet11} at VLT on 16 and 17 June 2013. Three stars (\#5, \#12, and \#67) are in common with the NIRSpec sample. We selected the VIS (slit=0.9”) and the NIR (slit=0.6”) arms to obtain nodded spectra at R$\approx$9,000 in the Calcium triplet(CaT) and NIR spectral ranges. Spectra in the CaT region were mostly used to compute the stellar RVs. 

\begin{figure*}
    \centering
    \includegraphics[width=0.95\textwidth]{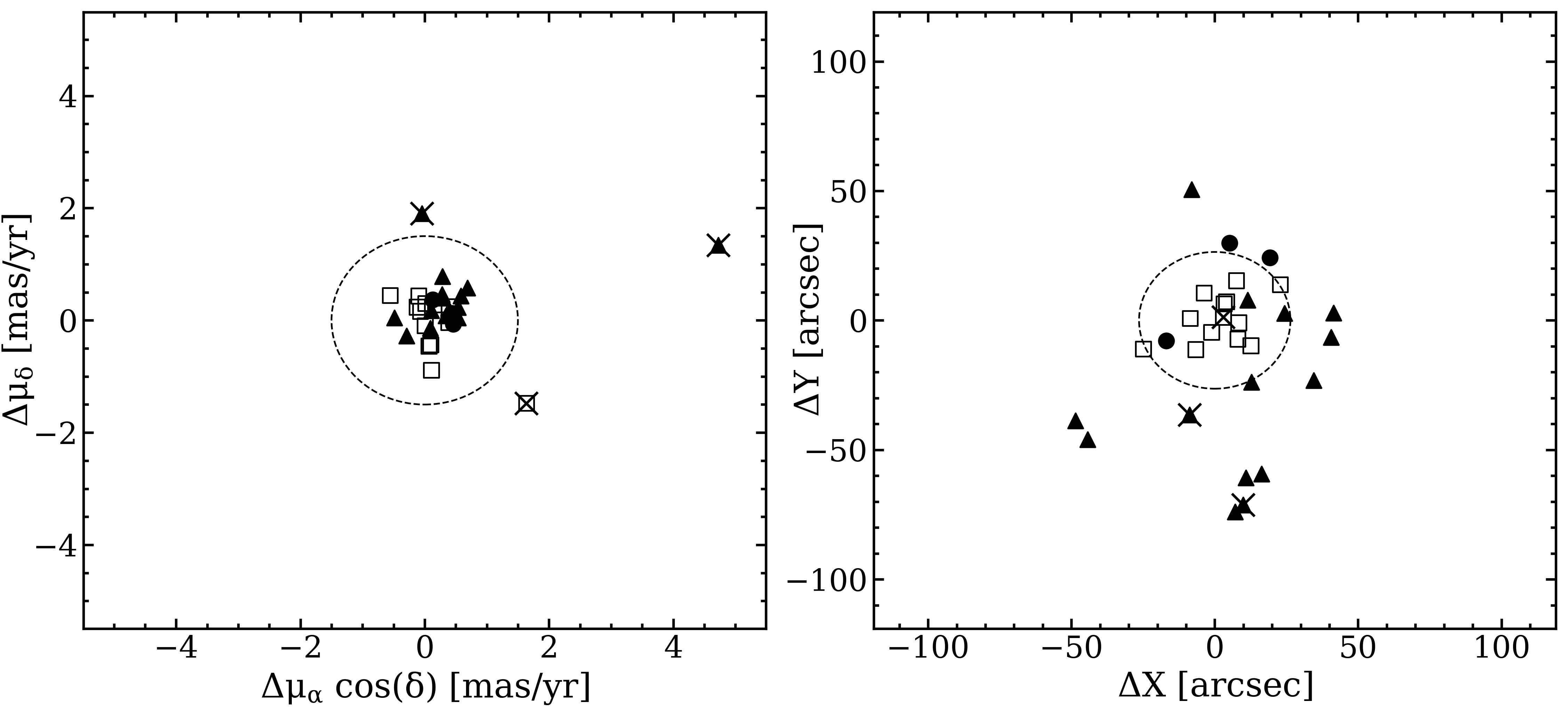}
    \caption{Kinematic and spatial location of the observed stars. Left panel: Vector point diagram of the 30 observed stars showing the $\alpha$ and $\delta$ components of the measured HST PMs referred to the systemic value ($\mu_{\alpha}\text{cos}\delta = -4.979$ mas yr$^{-1}$, $\mu_{\delta} = -7.431$ mas yr$^{-1}$) quoted by \citet{vasiliev_21}. The large dashed circle is centered on (0,0) and has a radius equal to $3\times\sigma_{\rm PM}$, with $\sigma_{\rm PM}=0.5$ mas yr$^{-1}$ being the measured dispersion of the PM distribution for the Terzan~6 bright stars. The three stars beyond this circle (marked with a cross) are considered as field interlopers. Right panel: Distribution of the observed stars in the plane of the sky with respect to the cluster center (same symbols as in the left panel). 
    A consistent fraction of them lie within the dashed circle marking the cluster half-light radius of 26.4\arcsec, while the remaining stars lie well within the tidal radius \citep[see, e.g.,][]{harris96}.}
    \label{vpd_radec}
\end{figure*}

Acquisition of both the NIRSpec and X-shooter spectra was performed by nodding on the slit with a typical throw of a few arcseconds for a proper background and detector subtraction. Each spectrum has been sky subtracted by using nod pairs, corrected for flat field, and calibrated in wavelength using arc lamps. 

An O-star spectrum observed during the same night was used to check and eventually remove telluric features. Typical total on-source exposure times ranged from 10 to 40 min.
The full reduction of the H-band NIRSpec spectra was performed by using the REDSPEC IDL-based package developed at the UCLA IR Laboratory. The reduction of the X-shooter 1.1-2.4 $\mu$m spectra was performed by using the ESO X-shooter pipeline to obtain 2D rectified and wavelength calibrated spectra. Order and 1D spectrum extraction were performed manually. The signal to noise ratio of the final spectra is always greater than 30. Figure~\ref{obs_stars} (right panel) shows portions of the H-band spectra for a giant star observed with both NIRSpec and X-shooter. While the higher spectral resolution of NIRSpec allowed for a more precise line measurement, the broader, simultaneous spectral coverage of the J-, H-, and K-bands of X-shooter allowed more diagnostic lines to be picked up, thus compensating for the lower resolution and ultimately providing chemical abundances with a similar precision.
Table~\ref{tab1} lists coordinates and NIR photometry from \citet{valenti07} of the observed stars of Terzan~6.

\section{Stellar kinematics and cluster membership}
\label{memb}
The cluster membership of the spectroscopic targets were checked by using relative PMs measured from the analysis of multi-epoch HST observations. The detailed analysis will be presented in a forthcoming paper (Loriga et al., in prep.). For the present study, we used ACS/WFC images of Terzan~6 in the optical bands acquired in two different epochs separated by a temporal baseline of $\sim 5$ yr. The PMs were measured by following the approach adopted in \citet{dalessandro13} and \citet{saracino19}. The procedure consists of determining the centroid displacement of the stars measured in different epochs with respect to a distortion-free \citep[see][]{anderson10,ubeda12} coordinate reference frame. The reference frame was then transformed into the absolute coordinate ($\alpha,\delta$) system by using the stars in common with the Gaia Early Data Release 3 catalog \citep{gaia_edr3}.  
The $\alpha$ and $\delta$ PM distribution of the red giant branch stars of Terzan~6 with luminosity comparable to that of the spectroscopic targets shows a dispersion $\sigma_{\rm PM}=0.5$ mas yr$^{-1}$ in both components. Hence, we selected as cluster members all the spectroscopic targets included within a circle of radius $3\times\sigma_{\rm PM}$ centered on the systemic PM of the cluster ($\mu_{\alpha}\text{cos}\delta = -4.979$, $\mu_{\delta} = -7.431$ mas yr$^{-1}$; \citealt{vasiliev_21}). 
Figure~\ref{vpd_radec} (left panel) shows the distribution of the spectroscopic targets in the vector point diagram, revealing that 27 out of the 30 observed stars are likely cluster members, while those with ID = 32, 33, and 74 are field interlopers. Figure~\ref{vpd_radec} (right panel) shows the distribution of the measured stars in the plane of the sky with respect to the 
new center of gravity determined by Loriga et al. (in prep.), which is located at RA=$267.6954167$ and Dec=$-31.2750389$ in a position that is approximately $8\arcsec$ away from the one quoted in \citet{harris96}. Interestingly, one of the three field interlopers is located (in projection) in the very central region of the cluster, thus further emphasizing the importance of a PM-based membership selection. The likely member stars are all included within just $\sim 80\arcsec$ from the cluster center.
For these stars, we measured the heliocentric RV with a  $\approx$1~km~s$^{-1}$ precision (see Table~\ref{tab2}). We found an average value of 143.3$\pm$1.0 km s$^{-1}$ and a dispersion $\sigma$ of 5.1$\pm$0.7 km s$^{-1}$. These values are in reasonable agreement with the systemic velocity (136.6$\pm$1.6 km s$^{-1}$) and central velocity dispersion (5.9$\pm$1.1 km s$^{-1}$) quoted in \citet{baumgardt18}; the latter, however, is based on a significantly smaller (almost by a factor of two) number of stars. The inferred RVs are all within $\pm 2\sigma$ from the average one, thus confirming the cluster membership from PMs of the selected targets.
\begin{table*}
\caption{Stellar parameters and chemical abundances for the observed stars in Terzan~6.}
\label{tab2}
\scriptsize
\setlength{\tabcolsep}{3.1pt}
\renewcommand{\arraystretch}{1.35}
\begin{tabular}{|c|c|c|c|c|c|c|c|c|c|c|c|c|c|c|c|}
\hline\hline
ID & T$_{\text{eff}}$ & log~g & RV & [Fe/H] & [Ca/H] & [Si/H] & [Mg/H] & [Ti/H] & [O/H] & [Al/H] & [Na/H] & [Mn/H] & [K/H] & [C/H] $^*$ & $^{12}$C/$^{13}$C $^*$\\
\hline
           5 &3500&0.5 &139&  -0.65 (.05)& -0.37 (.08) &-0.39 (.03) &-0.44 (.02) &-0.37 (.03) &-0.30 (.03) &-0.32 (.08) &-0.36 (.04) &-0.78 (.10) &-0.57 (.06) &-0.80 & 6\\
           9 &3500&0.5 &138&  -0.63 (.03)& -0.35 (.03) &-0.32 (.09) &-0.29 (.09) &-0.40 (.06) &-0.23 (.06) &-0.25 (.04) &-0.35 (.04) &-0.90 (.10) &-0.52 (.09) &-0.80 & 6\\
          10 &3600&0.5 &148&  -0.63 (.04)& -0.43 (.06) &-0.37 (.08) &-0.33 (.03) &-0.44 (.01) &-0.39 (.03) &-0.16 (.01) &-0.38 (.05) &-0.99 (.10) &-0.54 (.10) &-0.80 & 8\\
           12 &3600&0.5 &140&  -0.65 (.05)& -0.39 (.02) &-0.20 (.05) &-0.35 (.04) &-0.43 (.09) &-0.41 (.03) &-0.30 (.04) &-0.23 (.09) &-0.92 (.10) &-0.50 (.08) &-1.20 & 5\\
          13 &3600&0.5 &153&  -0.61 (.07)& -0.33 (.06) &-0.23 (.12) &-0.30 (.07) &-0.38 (.04) &-0.30 (.04) &-0.26 (.09) &-0.47 (.05) &-0.90 (.10) &-0.39 (.10) &-0.80 & 8\\
          14 &3600&0.5 &148&  -0.66 (.02)& -0.39 (.02) &-0.33 (.06) &-0.37 (.07) &-0.37 (.06) &-0.38 (.05) &-0.36 (.04) &-0.30 (.07) &-1.06 (.10) &-0.48 (.10) &-0.80 & 8\\
          18 &3600&0.5 &147&  -0.64 (.01)& -0.26 (.06) &-0.34 (.04) &-0.32 (.11) &-0.33 (.04) &-0.29 (.02) &-0.33 (.07) &-0.26 (.02) &-0.99 (.10) &-0.58 (.03) &-0.80 & 8\\
          20 &3800&0.5 &152&  -0.61 (.05)& -0.34 (.03) &-0.31 (.10) &-0.33 (.04) &-0.41 (.06) &-0.32 (.04) &-0.34 (.10) &-0.11 (.10) &-0.97 (.10) &-0.57 (.08) &-1.10 &10\\
          21 &3700&0.5 &154&  -0.67 (.04)& -0.30 (.05) &-0.33 (.11) &-0.32 (.02) &-0.45 (.07) &-0.29 (.02) &-0.40 (.06) &-0.31 (.09) &-0.98 (.10) &-0.51 (.07) &-0.80 & 9\\
          22 &3700&0.5 &140&  -0.67 (.04)& -0.35 (.07) &-0.28 (.10) &-0.29 (.01) &-0.32 (.08) &-0.35 (.07) &-0.39 (.09) &--  --      &--  --      &--  --      &-1.20 & 5\\
          28 &3700&0.5 &140&  -0.63 (.06)& -0.31 (.06) &-0.25 (.10) &-0.20 (.10) &-0.29 (.10) &-0.39 (.04) &-0.35 (.10) &--  --      &--  --      &--  --      &-1.10 & 5\\
          31 &3800&0.5 &137&  -0.65 (.02)& -0.36 (.04) &-0.25 (.04) &-0.27 (.11) &-0.38 (.05) &-0.33 (.06) &-0.32 (.03) &-0.17 (.13) &-1.00 (.10) &-0.59 (.10) &-1.15 & 5\\
          35 &3800&0.5 &140&  -0.67 (.04)& -0.31 (.05) &-0.27 (.10) &-0.36 (.05) &-0.31 (.07) &-0.29 (.06) &-0.24 (.04) &--  --      &--  --      &--  --      &-0.90 & 6\\
          37 &3800&0.5 &140&  -0.70 (.02)& -0.32 (.06) &-0.33 (.10) &-0.34 (.01) &-0.34 (.07) &-0.35 (.06) &-0.32 (.04) &--  --      &--  --      &--  --      &-1.20 & 6\\
          39 &3900&0.5 &140&  -0.69 (.09)& -0.28 (.09) &-0.37 (.10) &-0.33 (.08) &-0.34 (.01) &-0.32 (.05) &-0.26 (.03) &--  --      &--  --      &--  --      &-0.85 & 8\\
          46 &4000&1.0 &140&  -0.69 (.10)& -0.35 (.05) &-0.33 (.10) &-0.33 (.02) &-0.38 (.04) &-0.35 (.04) &-0.32 (.10) &--  --      &--  --      &--  --      &-0.80 & 6\\
          48 &3900&1.0 &153&  -0.65 (.03)& -0.26 (.04) &-0.27 (.06) &-0.22 (.03) &-0.33 (.03) &-0.44 (.04) &-0.24 (.05) &-0.08 (.16) &-0.85 (.10) &-0.52 (.07) &-1.10 & 7\\
          65 &4000&1.0 &140&  -0.62 (.08)& -0.26 (.06) &-0.23 (.10) &-0.22 (.03) &-0.23 (.10) &-0.25 (.05) &-0.23 (.10) &--  --      &--  --      &--  --      &-1.15 & 5\\
          67 &3900&1.0 &143&  -0.62 (.04)& -0.22 (.02) &-0.20 (.07) &-0.27 (.03) &-0.21 (.03) &-0.23 (.06) &-0.20 (.06) &-0.26 (.05) &-0.83 (.10) &-0.51 (.04) &-0.80 & 5\\
          69 &4250&1.5 &140&  -0.70 (.10)& -0.44 (.07) &-0.47 (.10) &-0.48 (.01) &-0.46 (.10) &-0.39 (.06) &-0.40 (.10) &--  --      &--  --      &--  --      &-0.80 & 6\\
          70 &4000&1.0 &144&  -0.62 (.06)& -0.34 (.09) &-0.34 (.03) &-0.35 (.06) &-0.16 (.11) &-0.32 (.06) &-0.23 (.06) &-0.11 (.07) &-0.89 (.10) &-0.49 (.04) &-1.00 & 8\\
          82 &4000&1.0 &144&  -0.63 (.04)& -0.30 (.05) &-0.25 (.11) &-0.31 (.11) &-0.29 (.06) &-0.30 (.05) &-0.28 (.07) &-0.20 (.16) &-0.85 (.10) &-0.47 (.03) &-1.00 & 8\\
          83 &4000&1.0 &140&  -0.65 (.01)& -0.23 (.10) &-0.23 (.10) &-0.29 (.02) &-0.24 (.03) &-0.23 (.02) &-0.26 (.10) &--  --      &--  --      &--  --      &-0.80 & 9\\
          90 &4000&1.0 &140&  -0.60 (.03)& -0.25 (.04) &-0.25 (.10) &-0.27 (.07) &-0.33 (.02) &-0.29 (.05) &-0.24 (.10) &--  --      &--  --      &--  --      &-0.90 & 9\\
         101 &4000&1.0 &140&  -0.65 (.02)& -0.35 (.06) &-0.39 (.10) &-0.26 (.12) &-0.34 (.13) &-0.38 (.03) &-0.34 (.07) &--  --      &--  --      &--  --      &-0.80 & 9\\
         114 &4250&1.5 &140&  -0.59 (.09)& -0.15 (.11) &-0.19 (.10) &-0.17 (.03) &-0.15 (.01) &-0.11 (.04) &-0.26 (.10) &--  --      &--  --      &--  --      &-0.85 & 6\\
         116 &4000&1.0 &139&  -0.68 (.05)& -0.31 (.02) &-0.26 (.05) &-0.41 (.12) &-0.37 (.10) &-0.41 (.06) &-0.40 (.12) &-0.21 (.12) &-0.97 (.10) &-0.66 (.07) &-0.80 & 8\\
         \hline
         \multicolumn{4}{|l|}{From GS98 to M22 solar reference}&0.00&-0.01&-0.04&+0.03&+0.08&+0.06&+0.04&+0.04&-0.13&-0.02&-0.04& -- \\
\hline\hline
\end{tabular}

\vspace{2pt}
$^*$ Carbon abundances from spectral synthesis. The typical error on [C/H] is $\pm$0.1 dex, and on the isotopic ratio $^{12}$C/$^{13}$C, it is $\pm$1.
\vspace{0.7cm}
\end{table*}

\section{Stellar parameters} 
\label{param}
The first estimates of the stellar temperature were derived from {the J-K} colors by using {the photometry and} the reddening estimates by \citet{valenti07} and the color-temperature scale by \citet{montegriffo98}, which was calibrated on GC giants. The gravity was estimated from theoretical isochrones \citep[e.g.,][]{bressan_12} {at a  metallicity of $-$0.7 dex, consistent with the photometric estimate by \citet{valenti07} and recently confirmed to be consistent with the spectroscopic one obtained in this study}. An average microturbulence velocity of 2 km s$^{-1}$, which provided a good fit to the observed features \citep[see, e.g.,][for a detailed discussion]{origlia97}, was adopted for all the stars.
The simultaneous spectral fitting of the CO and OH molecular lines that are especially sensitive to temperature, gravity, and microturbulence variations \citep[see also][]{origlia02} allowed us to fine-tune our best-fit adopted stellar parameters. Moreover, for the stars observed with X-shooter, we were also able to derive independent estimates of the temperature from the first-overtone (2-0) and (3-1) $^{12}$CO band heads in the K-band by using the calibration of \citet{schultheis16}.\\
For these stellar parameters, we obtained estimated uncertainties of $\pm$100K in T$_{\rm eff}$, $\pm$0.3 dex in log~g, and $\pm$0.3 km s$^{-1}$ in microturbulence, 
impacting the overall abundances at a level of 0.1-0.2 dex. Variations of temperature, gravity, and microturbulence smaller than the values quoted above are difficult to disentangle because of the limited sensitivity of the lines and the degeneracy among stellar parameters themselves. Moreover, such increased fine-tuning would have a negligible impact on the inferred abundances (less than a few hundredths of a dex; i.e., smaller than the measurement errors). Table~\ref{tab2} lists the final adopted temperatures and gravities.

\section{Chemical abundances}
\label{abun}
\begin{figure*}
    \centering
    \includegraphics[width=1\textwidth]{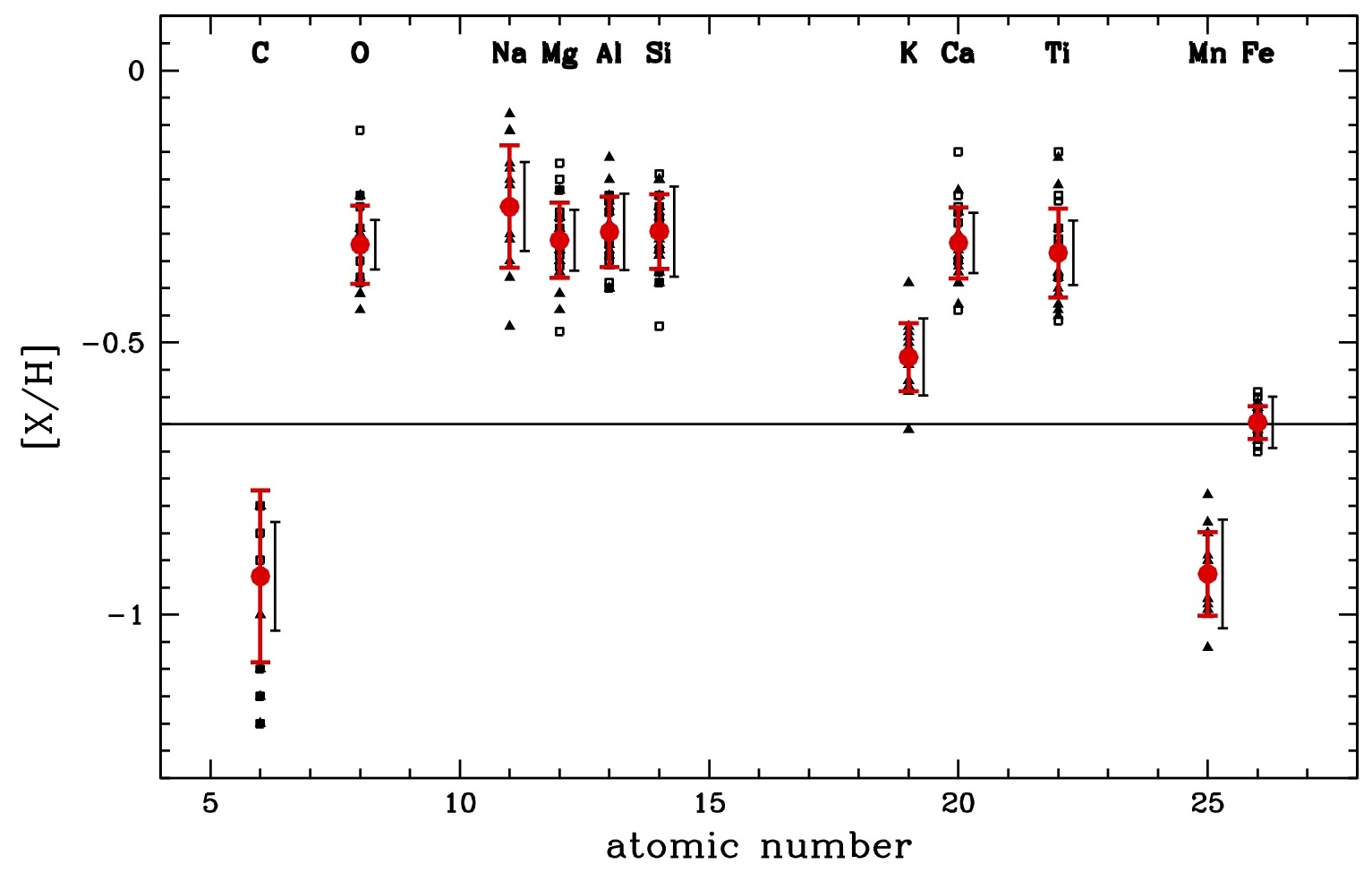}
    \caption{Chemical abundances for the Terzan~6 stars observed with X-shooter (filled triangles) and NIRSpec (open squares). Average values and corresponding 1$\sigma$ dispersions are overplotted in red, while typical measurement errors (black error bars) are also reported for comparison.}
    \label{abun}
\end{figure*}

\begin{figure*}
    \centering
    \includegraphics[width=0.98\textwidth]{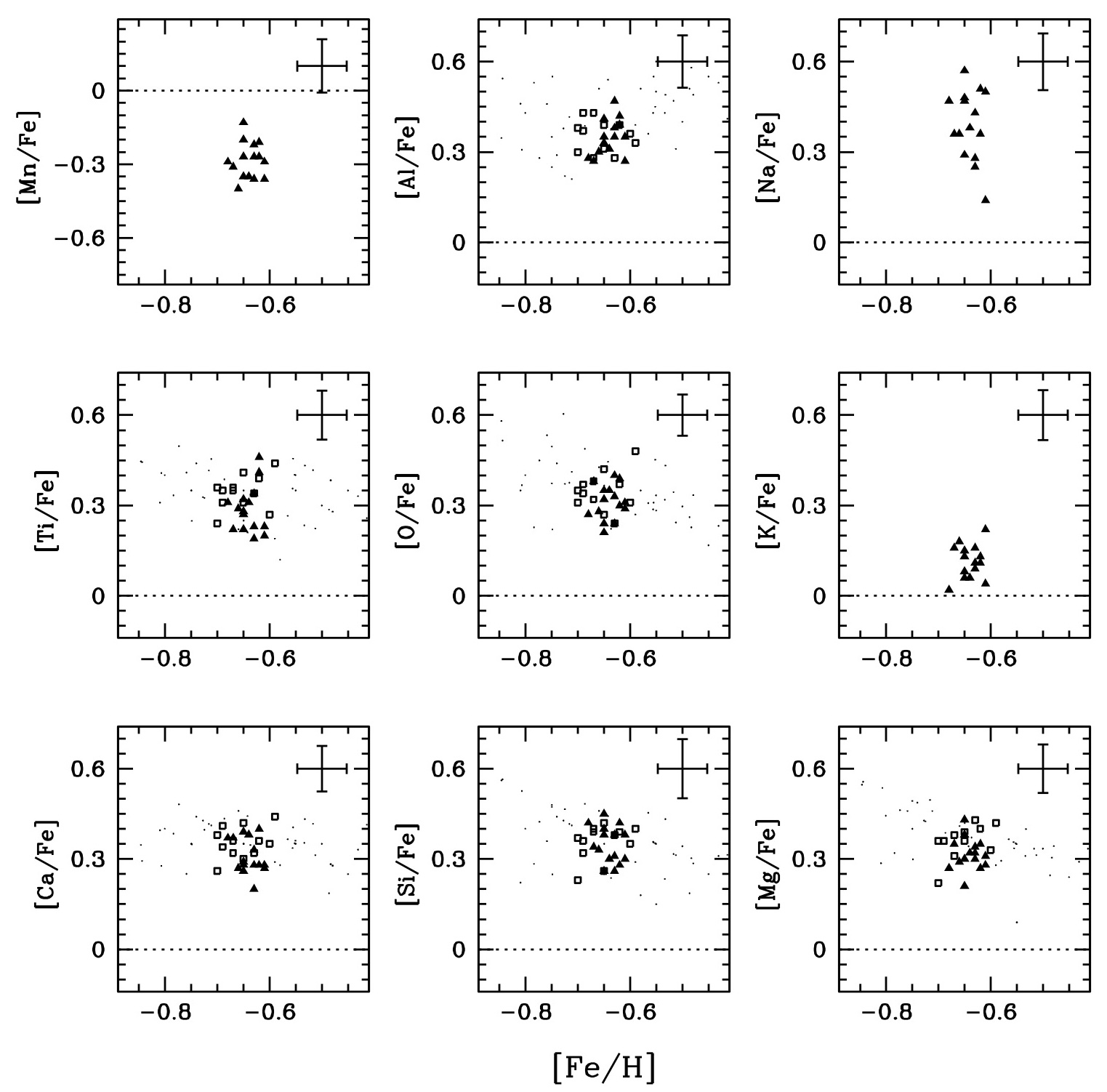}
    \caption{Chemical abundance patterns for the Terzan~6 stars observed with X-shooter (filled triangles) and NIRSpec (open squares). Typical error bars for the [Fe/H] and [X/Fe] abundance ratio values are also plotted in the corner of each panel. For comparison, the corresponding measurements of giant stars (dots) in other bulge GCs \citep[see][and references therein]{origlia08,valenti11} are also reported.}
    \label{ratio}
\end{figure*}

\begin{figure*}
    \centering
    \includegraphics[width=0.95\textwidth]{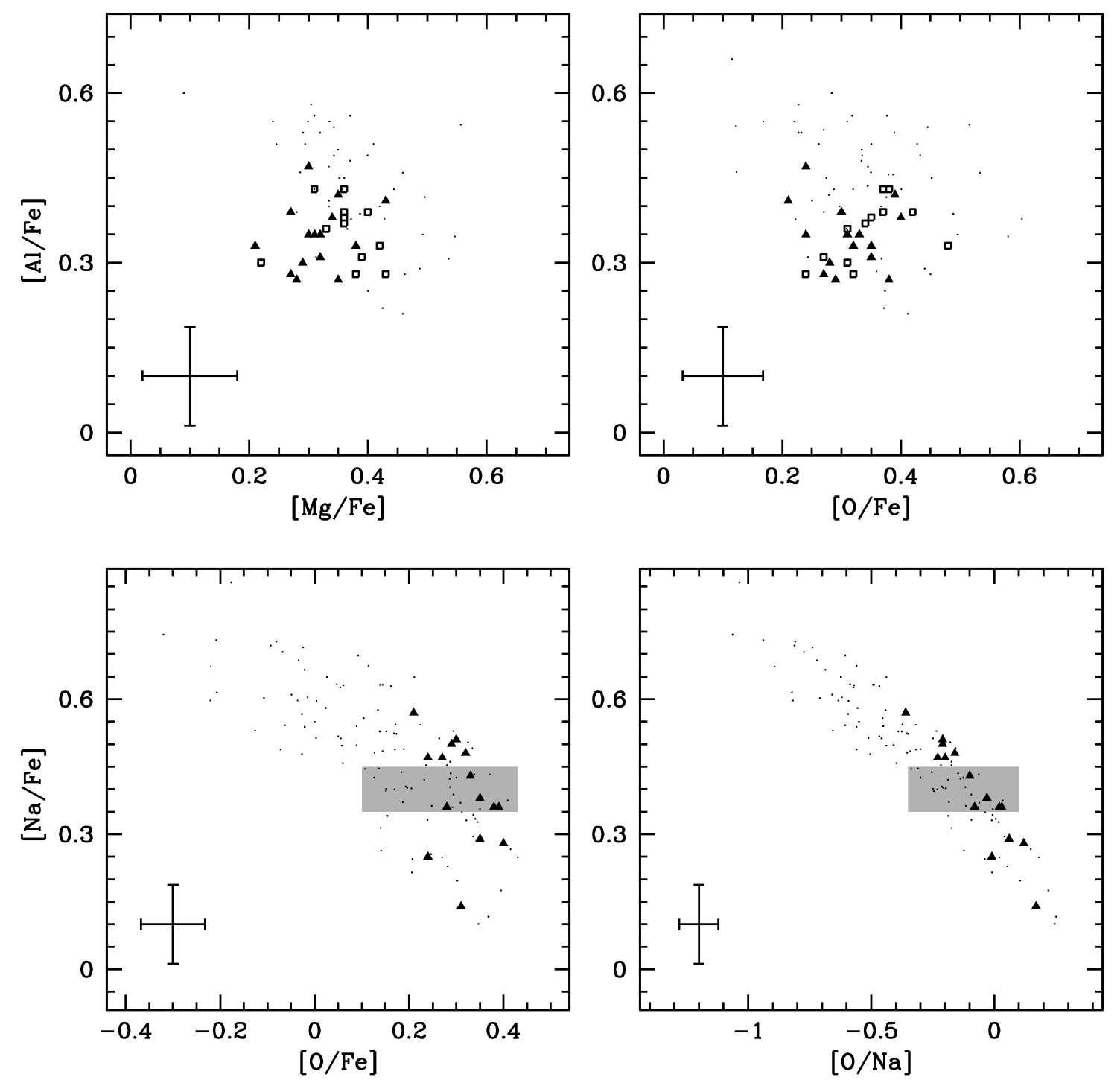}
    \caption{Abundance ratios between specific light elements. Bottom panels: [Na/Fe] versus [O/Fe] (left panel) and [O/Na] (right panel) abundance ratio distributions for the Terzan~6 stars observed with X-shooter (filled triangles). Typical measurement errors are also plotted in the corner of each panel. For comparison, the corresponding measurements of giant stars (dots) in 47~Tuc, a GC with similar metallicity, from the compilation of \citet{carretta09} are also reported. Stars below and above the gray shaded area are likely first and second generations, respectively, while those within it are borderline \citep[see also][]{carretta09}.
    Top panels: [Al/Fe] versus [Mg/Fe] (left panel) and [O/Fe] (right panel) abundance ratio distributions for the Terzan~6 stars observed with X-shooter (filled triangles) and NIRSpec (open squares). Typical  measurement errors are also plotted in the corner of each panel. For comparison, the corresponding measurements of giant stars (dots) in other bulge GCs \citep[see][and references therein]{origlia08,valenti11} and in 47~Tuc from the compilation of \citet{carretta09} are also reported.}
    \label{anti}
\end{figure*}

Chemical abundances of several metals were obtained by comparing the observed spectra with a suitable grid of synthetic ones covering the range of stellar parameters appropriate to the observed stars. These synthetic spectra were computed with the code described in detail in \citet{origlia02,origlia04}, which makes use of the local thermodynamic equilibrium approximation and the MARCS model atmospheres. It also includes thousands of NIR atomic transitions from the Kurucz database,\footnote{\url{http://www.cfa.harvard.edu/amp/ampdata/kurucz23/sekur.html}} while the molecular data were taken from our code \citep[][and subsequent updates]{origlia93} and B. Plez\footnote{\url{https://www.lupm.in2p3.fr/users/plez/}} compilations. We used the \citet[][hereafter, GS98]{Grevesse98} abundances for the solar reference, for homogeneity with our previous studies of bulge GCs. However, for completeness, in Table~\ref{tab2} we also provide the conversion factors from the GS98 solar reference system to the most recent one by \citet[][hereafter, M22]{magg22}.

A list of suitable lines for each measurable chemical species free from significant blending and/or contamination by telluric absorption and without strong wings has been identified both in the NIRSpec H-band spectra \citep[see also][]{origlia02,origlia04,origlia05} and in the JHK X-shooter (see, e.g., Fig.~1 of \citealt{deimer24}) ones.
Chemical abundances were derived by minimizing the scatter between the observed and synthetic spectra and by using as a figure of merit the line depths and equivalent width measurements and the overall spectral synthesis around each feature of interest.
Molecular lines of OH and neutral atomic lines of other light and heavy metals in the H-band NIRSpec and X-shooter spectra were used to derive abundances of O, Fe, Ca, Si, Mg, Ti and Al. Additional atomic lines of Mg, Al, K, and Mn in the J-band and of Ti, Al, and Na in the K-band X-shooter spectra were also used to derive abundances for the corresponding metals. 

The typical random errors in the measured line depths and equivalent widths mostly arose from a $\pm$1-2\% uncertainty in the placement of the pseudo-continuum, as estimated by overlapping the synthetic and the observed spectra. This error corresponds to abundance variations ranging from a few hundredths to one-tenth of a dex, and it is lower than the typical line-to-line abundance scatter ($\sim$0.15 dex). For each estimated element abundance, the error quoted in Tab.~\ref{tab2} was obtained by dividing the corresponding dispersion around the mean abundance value by the squared root of the number of used lines, typically a few per species. When only one line was available, we assumed a 0.1 dex error.
Molecular band heads of $^{12}$CO in the H-band NIRSpec and X-shooter spectra were used to derive $^{12}$C abundances. The $^{13}$C  carbon abundance was derived from the $^{13}$CO(4-1) band head in the H-band NIRSpec spectra and from the $^{13}$CO(2-0) band head in the K-band X-shooter spectra since the second overtone band heads in the H-band are too shallow at the lower spectral resolution of X-shooter.

The derived [X/H] chemical abundances of Fe, Ca, Si, Mg, Ti, O, Al, Na, Mn, K, and C and the isotopic abundance ratio $^{12}$C/$^{13}$C are listed in Tab. \ref{tab2} together with their measurement errors. 
For the three stars observed with both NIRSpec and X-shooter, we derived independent abundances for the elements in common, and since the two sets of values turned out to be fully consistent with each other (within the errors), we averaged them.
For each measured chemical element, Fig.~\ref{abun} shows the inferred abundances for the observed stars along with their mean value, 1$\sigma$ dispersion, and average measurement error, for comparison.
An average iron abundance of [Fe/H]=-0.65$\pm$0.01 dex was measured, which is fully consistent with previous estimates from the CaT \citep{az88} and NIR photometry \citep{valenti07}. 
The 1$\sigma$ dispersion, computed either as the r.m.s. of the observed distribution or by means of a maximum-likelihood method, turned out to be 0.03 dex, which is smaller than the average measurement error, thus indicating that Terzan~6 does not have any appreciable intrinsic spread in iron. 
We note that Ca, Si, Mg, Ti, O, Al, Na, and K have [X/H] abundances exceeding the corresponding solar-scaled values, while C and Mn are depleted. The 1$\sigma$ dispersions exceed the corresponding average measurement error in the cases of C, O, Ti, Na, and, to a lower extent, of Mg, while no significant dispersion within the measurement error has been found in the cases of Al, K, and Mn.

Figure~\ref{ratio} shows the measured [Ca/Fe], [Si/Fe], [Mg/Fe], [Ti/Fe], [O/Fe], [Al/Fe], and [Na/Fe] abundance ratios as a function of [Fe/H]. All of these abundance ratios are enhanced with respect to the corresponding solar values by 0.3-0.4 dex, while [K/Fe] is only mildly enhanced by an average of 0.11 dex.
For all of these abundance ratios but [Na/Fe], a 1$\sigma$ dispersion of $<$0.1 dex was derived, consistent with the measurement errors. The 1$\sigma$ dispersion (0.12 dex) of [Na/Fe] moderately exceeds (by about 23\%) the average measurement error. 
Figure~\ref{ratio} also shows an overall [Mn/Fe] depletion with respect to the solar value by a factor of two and a low spread within the measurement errors, in agreement with the values measured in other cluster and field giants of similar metallicity \citep[see, e.g.,][and references therein]{carretta04,sobeck06,barbuy13,schultheis17} as well as with model predictions for the bulge by \citet{cescutti08}.
An average [C/Fe]=-0.28 dex carbon depletion with a 1$\sigma$ dispersion of 0.16 dex and an average low $^{12}$C/$^{13}$C isotopic ratio of 7.1 with a 1$\sigma$ dispersion of 1.5 were inferred, which is similar to what has been measured in other bulge field and GC giants \citep[see, e.g.,][]{origlia08,rich12} and consistent with the occurrence of some mixing and extra mixing processes in the stellar interiors during the post main sequence evolution \citep[see, e.g.,][]{cha95,dw96,csb98,bs99} and in the high metallicity regime.

Figure~\ref{anti} shows the behavior of specific abundance ratios between light elements, namely [Na/Fe] as a function of [O/Fe] and [O/Na], and [Al/Fe] as a function of [Mg/Fe] and [O/Fe], that normally show some scatter in GCs \citep[see, e.g.,][]{carretta10}. 
We measured [Na/Fe] values ranging from +0.1 to +0.6 dex, [O/Na] values ranging from $-$0.4 to +0.2 dex, while [O/Fe], [Al/Fe], and [Mg/Fe] values span ranges that do not exceed 0.3 dex.  

\section{Discussion and conclusions}
\label{conc}

We have presented the first comprehensive chemical study of the Terzan~6 stellar population based on NIR medium-high resolution spectroscopy of a representative sample of giant stars. 
The inferred chemical abundances and abundance ratios and the associated dispersions of the measured stars are consistent with Terzan~6 being a genuine GC of the Galactic bulge (whose in situ origin has also been dynamically probed by, e.g., \citet{mas19}). Its average iron abundance of $-$0.65 dex matches well with the distribution of the metal-rich population of bulge GCs at [Fe/H]$>$-1.0 \citep[see, e.g.,][and references therein]{origlia08,valenti11,geisler21,kader22,fertrin22}. 

Its enhanced alpha element content is fully consistent with the old age of its stellar population, which likely formed at early epochs from a gas mainly enriched by type II supernovae, 
similar to the old populations of GCs in general, either in the halo or in the bulge, regardless of their metallicity.

Some scatter in the light element abundances is consistent with the possible self-enrichment of Terzan~6 in these elements during its early lifetime, as is typically observed in Galactic GCs and interpreted as a signature of multiple stellar populations \citep{carretta10}.
In particular, as shown in Fig.~\ref{anti}, the measured scatters in the [Na/Fe] and [O/Na] abundance ratios are consistent with those measured in first-generation and intermediate second-generation stars of GCs with a similar metallicity, such as 47 Tuc \citep[see, e.g.,][]{carretta09}. 
Indeed, in Terzan~6, we found four stars (likely first generation) with [O/Na]$<$0.3 dex and a roughly positive [O/Na], six stars (likely second generation) with [Na/Fe]$>$0.45 dex and [O/Na]$<$-0.1 dex, and five stars with somewhat borderline values of 0.35$<$[Na/Fe]$<$0.45 and $-$0.1$<$[O/Na]$<$0.1 dex. We can thus reasonably foresee a fraction between approximately 30\% and 50\% of first-generation stars, matching well with the typical values derived for other GCs with similar luminosities, including those in the bulge \citep[see, e.g.,][]{carretta09,kader22}. 
The lack of an extended anticorrelation between [Na/Fe] and [O/Fe]
and similarly between [Al/Fe] and [Mg/Fe] or [O/Fe](see Fig.~\ref{anti}) in Terzan~6, 
at variance with
what has been found in 47~Tuc, is however consistent with the former being significantly less massive than the latter and with has been found in other clusters of similar mass, luminosity, and metallicity \citep[see, e.g.,][]{carretta09,carretta10,caloi11,milone17,villanova19}. 

\begin{acknowledgements}
We thank the anonymous referee for his/her detailed report and useful comments and suggestions. CF and LO acknowledge the financial support by INAF within the VLT-MOONS project. DM acknowledges financial support from the European Union – NextGenerationEU RRF M4C2 1.1  n: 2022HY2NSX. "CHRONOS: adjusting the clock(s) to unveil the CHRONO-chemo-dynamical Structure of the Galaxy” (PI: S. Cassisi). This work is part of the project Cosmic-Lab at the Physics and Astronomy Department “A. Righi” of the Bologna University (http:// www.cosmic-lab.eu/ Cosmic-Lab/Home.html).
\end{acknowledgements}

\bibliographystyle{aa} 
\bibliography{mybib} 

\end{document}